\documentclass[english]{article}
\usepackage[T1]{fontenc}
\usepackage[latin9]{inputenc}
\usepackage{amssymb}
\usepackage{esint}
\usepackage{babel}
\begin{document}

\title{Metaphysics of the free Fock space with local and global information}

\author{Jerzy Han\'{c}kowiak}

\author{(former lecturer and research worker of Wroclaw and Zielona Gora
Universities)}

\author{Poland}

\author{EU}

\author{e-mail: hanckowiak@wp.pl}

\date{March, 2012}
\maketitle
\begin{abstract}
A new interpretation of the basic vector $|0>$ of the free Fock space
(FFS) and the FFS is proposed. The approximations to various equations
with additional parameters, for n-point information (n-pi), are also
considered in the case of non-polynomial nonlinearities.

\end{abstract}
\tableofcontents{}

\section{Introduction and basic axioms}

Can sometimes happen that scientists resemble somewhat the Moliere's
hero, Mr Jourdain, who did not know that he is speaking prose. I see
this analogy in fact in ignoring of the \textbf{free Fock space} (FFS)
as an arena in which description should be made of various complex
phenomena and processes like turbulence, economic or general relativity
quantum processes. My faith in possibilities of the FFS comes from
the fact that in this space you can describe both macro- and micro-physics,
see \cite{Han 2011}. Moreover, in these both cases one can introduce
vacuum vectors as well as the cration and annihilation operators and
develope similar calculus, \cite{Han 2010,Han 2011},\cite{Han 2011-2}.
It turns out that in the FFS all these complex phenomena can be formulated
in a linear way. Moreover, in the FFS in a natural way one side-invertible
operators appear by means of which such important concepts as injectivity
and surjectivity can be defined in a way that does not depend on elements
of a set. This, in some sense, suggests a connection of the FFS with
the category theory - promising land of combination of different concepts,
see \cite{Baez 1999},\cite{Adam 2004} and Appendix2.

The vector $|V>$ describing, e.g., a classical or quantum system
can be expanded in an infinite series of the linearly independent
orthonormal vectors $|\tilde{x}_{(n)}>$:

\begin{equation}
|V>=\sum_{n=1}\int d\tilde{x}_{(n)}V(\tilde{x}_{(n)})|\tilde{x}_{(n)}>+V(0)|0>\label{eq:1}
\end{equation}
 The vectors (\ref{eq:1}) form the \textbf{Fock space} when basic
vectors $|\tilde{x}_{(n)}>$ are constructed in a certain way with
the help of creation operators and the basic vector |0>, see below.
The adjective '\textbf{free}'\textbf{ before }'\textbf{Fock space}'\textbf{
}comes from the fact that the 'components' $V(\tilde{x}_{(n)})$ of
vector |V> are not restricted by any symmetries, such as permutation
invariance. Algebraic realization of this fact is given by creation
and annihilation operators satisfying the Cuntza relations, see below.
See also \cite{Han 2010-1}. In our opinion, the \textbf{algebraic
thread introduced by the FFS} is more fruitful and essentional than
approaches based on orthonormal vectors (\ref{eq:1}) or equivalent
approach based on the linear functionals:

\[
V=\sum_{n=1}\int d\tilde{x}_{(n)}V(\tilde{x}_{(n)})\rho(\tilde{x}_{(n)})
\]
with the 'arbitrary' functions $\rho(\tilde{x}_{(n)})\equiv\rho(\tilde{x}_{1},...,\tilde{x}_{n})$
and the corresponding functional derivatives: $\delta/\delta\rho(\tilde{x}_{1},...,\tilde{x}_{n})$,
\cite{Hank 1992}. Of course, all these representations are equivalent. 

The components of the vectors $\tilde{x}$, on which the components
$V(\tilde{x}_{(n)})$ of the vector |V> and the basic vectors $|\tilde{x}_{(n)}>$depend,
contain the time-space and other (discrete ) variables. Moreover,
for the most part the work, we would assume that the time-space coordinates
are discrete. However, for greater transparency in the work, we use
the symbol $\int$ instead of $\sum$. 

The adjective 'free' also refers to the fact that the vector variables
$\tilde{x}_{(n)}=\tilde{x}_{1},...,\tilde{x}_{n}$are independent
of each other, for example, their time components are not equal to
one another. In other words, under the sign of smoothing or averaging,
<>, we mean rather the tensor product of fields so that the \textit{multi-time}
n-pi are considered. In this way the concept of the \textbf{state
vector of the system} is substituted by the concept of the \textbf{generating
vector of the system}. It seems that this replacement is particularly
useful in the areas, where the averagings or smoothings are used,
see also \cite{Han 2011-2,Han 2011}, \cite{Han 2008}.

\section*{Interpretation of |V>}

The 'components' $V(\tilde{x}_{(n)})$ have physical interpretations,
e.g., they can be some summarizing, averaging or smoothing of product
of the field $\varphi$: $<\varphi(\tilde{x}_{1})\cdots\varphi(\tilde{x}_{n})>$
, see e.g., \cite{Han 2008}, \cite{Han 2011}, or App.1and App.4.
They are called the \textbf{\textit{n-point (quasilocal) information
(about the system or object), (n-pi)}}. 

The vector $|\tilde{x}_{(n)}>$ standing at the given n-pi represents
rather everything that leads to the determination (measurement) of
the n-pi $V(\tilde{x}_{(n)})$. The vector |V> thus refers to a physical
system, such as electromagnetic fields or the economic system described
by an infinite system of n-pi and associated to them infinite system
of measuring arrangements. Assuming the possibility of perfect isolation
from the rest of the universe, we can assume that anything beyond
the given system and used instrumentation does not exist. Since the
vector |0> does not represent any measuring instrument, and the factor
V (0) does not include any local information on the system, we can
assume that $V(0)|0>\equiv|0>_{info}$ represents a local vacuum.
At this point, we break the tyranny of the philosophy of eigenvectors
and eigenvalues of at least some unifying scientific descriptions.
See also \cite{baker 2008}. 

We accept the following assumptions (axioms) concerning certain vectors
belonging to the FFS \textbf{formed with vectors related to particular
or total information about the Universe}:

\section*{Axioms:}

ASSUMPTION1 (A1):

There is a vector $|U>$ representing the Universe (Total system,Total
(Global) information)

ASSUMTION2 (A2):

Vectors |V> like (\ref{eq:1}) represent systems belonging to the
Universe (subsystems, (quasi) global information)

ASSUMPTION3 (A3):

There is a vector $|0>_{info}$ representing the local vacuum (no
local information) such that:

\begin{equation}
|0>_{info}\sim|0>\sim|U>\label{eq:2}
\end{equation}
$\bullet$

\section*{Comments:}

The Universe is understood in the sense of category theory, see \cite{Adam 2004}
and Appendix2. 

The vector $|0>$ is a basic vector of the FFS by means of which the
orthonormal base vectors $|\tilde{x}_{(n)}>$, for $n=0,1,2,...,\infty$,
and dual to them $<\tilde{x}_{(m)}|$ , for $m=0,1,2,...,\infty$,
are constructed. We have

\begin{equation}
V(\tilde{x}_{(n)})=<\tilde{x}_{(n)}|V>\label{eq:3}
\end{equation}
 for n=1,2,.... and, for n=0, $V(0)=<0|V>$. 

In the FFS the base vectors $|\tilde{x}_{(n)}>$ are constructed using
the vector $|0>$ and the creation operators $\hat{\eta}^{\star}(\tilde{x})$
as follows: 

\begin{equation}
|\tilde{x}_{(n)}>=\hat{\eta}^{\star}(\tilde{x}_{1})\cdots\hat{\eta}^{\star}(\tilde{x}_{n})|0>\label{eq:4}
\end{equation}
 In other words, every measuring arrangements are 'constructed' from
the basic vector |0>. This probably means that in A3 you can identify
the vector |0> with the vector |U>. 

\begin{equation}
|U>=|0>\label{eq:5}
\end{equation}
 Moreover,  proportional identification of these two vectors with
a local vacuum $|0>_{info}$ expresses the possibility of isolation
of the given system from the rest of the Universe. The rest of the
Universe identified with the whole Universe works like a vacuum system. 

This is consistent with the intuitive understanding of the concepts
- local and global. Moreover, as we know, in various areas of material
and spiritual world, global phenomena have negligible or no impact
on local phenomena, and vice vers, and it seems surprisingly that
this property is confirmed by the derived equations. 

To better feel the interaction, or rather the lack of it, between
global and local phenomena in economics, see \cite{Thur 2003}.

\section*{Cuntz relations}

In algebraic language the FFS can be constructed by means of creation
and annihilation operators satisfying the Cuntz relations: 

\begin{equation}
\hat{\eta}(\tilde{y})\hat{\eta}^{\star}(\tilde{x})=\delta(\tilde{y}-\tilde{x})\cdot\hat{I}\label{eq:6}
\end{equation}
and

\begin{equation}
\hat{\eta}(\tilde{y})|0>=0,\quad<0|\hat{\eta}^{\star}(\tilde{x})=0\label{eq:7}
\end{equation}
where all variables $\tilde{x},\tilde{y}$ are discrete quantities
and $\hat{I}$ is the unit operator in the FFS. In this way $\delta$
denotes rather Kronecker not Dirac symbol. In a sense, the Cuntz relations
remind us of the relations: $\partial/\partial x_{i}\cdot x_{j}=\delta_{ij}$
, but from the Cuntz relations the analogue of the Leibniz identity
does not result, see \cite{Han 2011-2}. For simplicity, we assume
that the creation operators $\hat{\eta}^{\star}(\tilde{x})$ are Hermitian
conjugate to the appropriate annihilation operators $\hat{\eta}(\tilde{x})$
.

\section*{Additional remarks}

The components $V(\tilde{x}_{(n)});\: n=1,2,...,\infty$ of the vector
$|V>$, standing at the vectors $|\tilde{x}_{(n)}>$ in the expansion
(\ref{eq:1}) quantitatively and specifically describe separate parts
of the system at varies times, see, e.g., \cite{Han 2011}. In fact,
we consider here the multitime n-pi and in fact this is responsible
that such vectors as $|V>,|\tilde{x}_{(n)}>$ do not represent states! 

\textbf{See Final remarks} in order to clarify the content of the
work!

\section*{Illustration of axioms:}

\begin{eqnarray*}
 & \longrightarrow\longrightarrow\longrightarrow\longrightarrow\longrightarrow\rightarrow & A1\\
 & \longrightarrow\longrightarrow\longrightarrow & A2\\
 & \leftrightarrow & A3
\end{eqnarray*}

\section{Equations for the generating vector |V>}

The vector $|V>\in FFS$ can describe an oscillator or the Universe
in a depence on the choice of the linear operator $\hat{A}$ appearing
in the equation:

\begin{equation}
\hat{A}|V>=|0>_{info}\label{eq:8}
\end{equation}
 In the both cases, the\textit{ vacuum vector} $|0>_{info}\sim|0>$.
Eq.\ref{eq:8} restricting n-pi generated by the vector $|V>$ was
derived in previous papers \cite{Han 2011,Han 2011-2}, so we only
here remind you some its properties, for a more, see App1. It turns
out that the operator $\hat{A}$ can be decompose in three parts each
of which is related to a specifice interaction of the system and we
get the following equation:

\begin{equation}
\left(\hat{L}+\lambda\hat{N}+\hat{G}\right)|V>=\hat{P}_{0}\hat{L}|V>+\lambda\hat{P}_{0}\hat{N}|V>+\hat{P}_{0}\hat{G}|V>\equiv|0>_{info}\label{eq:9}
\end{equation}
where the vectot $|0>_{info}$ describes so called the local information
vacuum. In other words, no local information is included in this vector.
In this paper the vector $|0>_{info}$is called the \textit{classical
vacuum}.

The operator $\hat{L}$ is a right invertible operator, which is diagonal:

\begin{equation}
\hat{P}_{n}\hat{L}=\hat{L}\hat{P}_{n}\label{eq:10}
\end{equation}
 with respect to the projectors $\hat{P}_{n}$, where the project
$\hat{P}_{n}$projects on the n-th term in the expansion (\ref{eq:1}). 

The operator $\hat{G}$ is a left invertible operator, which is lower
triangular operator:

\begin{equation}
\hat{P}_{n}\hat{G}=\sum_{m<n}\hat{P}_{n}\hat{G}\hat{P}_{m}\label{eq:11}
\end{equation}
and the operator $\hat{N}$ can be right or left invertible operator,
see \cite{Han 2011-2} (all depend on a definition of the operator-valued
functions). In the case of polynomial operators this is an upper triangular
operator:

\begin{equation}
\hat{P}_{n}\hat{N}=\sum_{n<m}\hat{P}_{n}\hat{N}\hat{P}_{m}\label{eq:12}
\end{equation}
see \cite{Han 2011}. 

All these operators are linear in the FFS. 

The diagonal operator $\hat{L}$ is related to the linear interaction
among constituents of the system and, or, describes the kinematic
part of theory usually responsible for additional conditions needed
for the unique solutions. In the case of quantum field theory this
part of the operator $\hat{A}$ corresponds to lack of interaction
between the system components. If we take into account that $\hat{L}$
is a surjection (the image of the domain of a surjective function
completely covers the function's codomain) then it is surprising that
\textbf{Axiom of Choice} of set theories is equivalent to saying:
\textbf{any surjective function has a right inverse! }See App.2. 

The operator $\hat{N}$ is related to the nonlinear interaction of
the constituents including self-interaction as well. It is this property
which leads, independently of right or left invertability of the operator
$\hat{N}$, to an infinite system of equations to n-pi, see \cite{Han 2011-2,Han 2011}. 

The last, lower trangular operator $\hat{G}$ expresses very deep,
qualitative properties of described systems like this that the system
is immersed in a given external field, or that the system is subjecting
to certain constraints (commutation relations) that are implemented
without the participation of reaction forces, see (\cite{Han 2011};
Secs 3 and 4.2). In the case of \textbf{General Relativity} the operator
$\hat{G}$ is related to the stress-energy momentum tensors and in
this case should be rather denoted as 

\[
\hat{G}\equiv\hat{T}
\]
In the case of \textbf{Macroscopic Gravity} (MG) to descibe large
scale structures entering the Cosmos, the above operator can be treated
as a random quantity from which the n-pi (e.g.,correlation functions)
depend parametrically. After solving in formal way the equations for
the n-pi obtained by means of averaging with respect to initial and
boundary conditions, for Eq.\ref{eq:71}, we can calculate appropriate
expectation values with respect to $\hat{T}$. In this way the averaging
problem related to the space-time averaging can be avoided, see \cite{Wilt 2011}
and App.6. 

Taking into account that 

\begin{equation}
\hat{P}_{0}\hat{G}=0\label{eq:13}
\end{equation}
 we see that Eq.\ref{eq:9}, for projection $\hat{P}_{0}$, is identically
satisfied. However, \textbf{we have introduced such identity to have
in its result, in the l.h.s. of (\ref{eq:9}), where it is possible,
the right invertible operators}, see \cite{Han 2011,Han 2011-2}.
The term, which have had appeared in the r.h.s., we interpreted as
a (local) vacuum. In the present study we want to examine its role.

\section{A role of the classical vacuum represented by the vector $|0>_{info}$}

It turns out that complement of the equations for the generating vector
$|V>$ in such a way that the operators appearing in them were at
least one side invertible, caused the appearance in homogeneous equations
the vector $|0>_{info}$, which we called the (local) vacuum.

\subsection{$\hat{L}$ is a right invertible operator. The perturbation principle}

Taking into account that the operator $\hat{L}$ is right invertible
it means that a right inverse exists $\hat{L}_{R}^{-1}$, for which

\begin{equation}
\hat{L}\hat{L}_{R}^{-1}=\hat{I}\label{eq:14}
\end{equation}
 With the help of operator $\hat{L}_{R}^{-1}$ we can rewrite Eq.\ref{eq:9}
as follows

\begin{equation}
\left(\hat{I}+\lambda\hat{L}_{R}^{-1}\hat{N}+\hat{L}_{R}^{-1}\hat{G}\right)|V>=\hat{L}_{R}^{-1}|0>_{info}+\hat{P}_{L}|V>\label{eq:15}
\end{equation}
 where a projector

\begin{equation}
\hat{P}_{L}=\hat{I}-\hat{L}_{R}^{-1}\hat{L}\label{eq:16}
\end{equation}
 projects on the null space of the operator $\hat{L}$, see \cite{Przeworska 1988}.
It is a diagonal projector, see (\ref{eq:10}), and its choice, among
other things, depends on a choice of a right inverse operator $\hat{L}_{R}^{-1}$. 

We will assume that solutions are symmetric:

\begin{equation}
|V>=\hat{S}|V>\label{eq:17}
\end{equation}
for example - the permutation symmetric. Because the operator $\hat{L}_{R}^{-1}\hat{G}$
is a lower triangular operator and the operator $\hat{S}$ is diaginal,
see \cite{Han 2010-1} , the Eq.\ref{eq:15} can be equivalently transformed
further as follows:

\begin{eqnarray}
 & \left(\hat{I}+\lambda\left(\hat{I}+\hat{S}\hat{L}_{R}^{-1}\hat{G}\right)^{-1}\hat{S}\hat{L}_{R}^{-1}\hat{N}\right)|V>=\nonumber \\
 & \left(\hat{I}+\hat{S}\hat{L}_{R}^{-1}\hat{G}\right)^{-1}\left(\hat{S}\hat{L}_{R}^{-1}|0>_{info}+\hat{S}\hat{P}_{L}|V>\right)\label{eq:18}
\end{eqnarray}
 This equation allows the next (formal) step:

\begin{eqnarray}
 & |V>=\left(\hat{I}+\lambda\left(\hat{I}+\hat{S}\hat{L}_{R}^{-1}\hat{G}\right)^{-1}\hat{S}\hat{L}_{R}^{-1}\hat{N}\right)\bullet^{-1}\nonumber \\
 & \left(\hat{I}+\hat{S}\hat{L}_{R}^{-1}\hat{G}\right)^{-1}\left(\hat{S}\hat{L}_{R}^{-1}|0>_{info}+\hat{S}\hat{P}_{L}|V>\right)\label{eq:19}
\end{eqnarray}
 which can be used for further calculations. Note first that the expression
$\left(\hat{S}\hat{L}_{R}^{-1}|0>_{info}+\hat{S}\hat{P}_{L}|V>\right)$
is not yet known. The vector $\hat{S}\hat{P}_{L}|V>$, arbitrary from
the standpoint of Eq.\ref{eq:9}, we fix by taking into account the
\textit{perturbation principle} according to which 

\begin{equation}
\hat{S}\hat{P}_{L}|V>=\hat{S}\hat{P}_{L}|V>^{(0)}\label{eq:20}
\end{equation}
where superscript (0) means that $\lambda=0$ in the l.h.s. of Eq.\ref{eq:17},
see, however, App.7. 

We still do not know the vacuum vector $|0>_{info},$ see Eq.\ref{eq:9}.
We assume the following \textit{normalization condition}:

\begin{equation}
\hat{P}_{0}\hat{L}|V>=|0>\label{eq:21}
\end{equation}
 and from Eq.\ref{eq:16} we have upon the term $\hat{P}_{0}\hat{N}|V>$the
following equation:

\begin{equation}
\hat{P}_{0}\hat{N}|V>=\hat{P}_{0}\hat{N}\bullet\left(r.h.s.\right)\: of\:(19))\label{eq:22}
\end{equation}
 or equivalently:

\begin{equation}
<0|\hat{N}|V>=<0|\hat{N}\bullet\left(r.h.s.\right)\: of\:(19)\label{eq:23}
\end{equation}
 This is the equation for the element $<0|\hat{N}|V>$, which also
occurs in the r.h.s. of Eq.\ref{eq:19}, see Eq.\ref{eq:9}. 

Denoting bra-vector $<0|\hat{N}\equiv<\Psi|$ we postulate the following
equation on $<\Psi|$:

\begin{equation}
\frac{\delta|V>}{\delta<\Psi|}=0\label{eq:24}
\end{equation}
 This equation comes from the fact that the extension of the operator
$\hat{N}$ about the element $\hat{P}_{0}\hat{N}$ is in a sense,
arbitrary. Hence the desire to minimalize with respect to changes
of $<\Psi|$ computed quantities .

\subsection{A theory with two parameters (coupling constants) and left invertible
$\hat{N}(\lambda_{2})$}

In \cite{Han 2010}, see also \cite{Han 2008}, we considered a theory
with two parameters $\lambda_{1}$ and $\lambda_{2}$with the following
equation on the generating vector |V>:

\begin{eqnarray}
 & \left(\hat{L}+\lambda_{1}\hat{N}(\lambda_{2})+\hat{G}\right)|V>=\nonumber \\
 & \hat{P}_{0}\left(\hat{L}+\lambda_{1}\hat{N}(\lambda_{2})+\hat{G}\right)|V>\equiv|0>_{info}\label{eq:25}
\end{eqnarray}
 In this case the parameter $\lambda_{1}$is called the \textit{major
coupling constant}. We additionally assume that the operator $\hat{N}$
is a left invertible, see (\ref{eq:44}). It means that an operator
$\hat{N}_{l}^{-1}(\lambda_{2})$ exists, for which

\begin{equation}
\hat{N}_{l}^{-1}(\lambda_{2})\hat{N}(\lambda_{2})=\hat{I}\label{eq:26}
\end{equation}
see \cite{Han 2011-2} and Eq.\ref{eq:47}in the present paper. Hence
and from Eq.\ref{eq:25} we get

\begin{equation}
\left(\hat{N}_{l}^{-1}(\lambda_{2})(\hat{L}+\hat{G})+\lambda_{1}\hat{I}\right)|V>=\hat{N}_{l}^{-1}(\lambda_{2})|0>_{info}=0\label{eq:27}
\end{equation}
 Taking into account that the operator $\hat{N}_{l}^{-1}(\lambda_{2})(\hat{L}+\hat{G})$
is a right invertible with operator $(\hat{L}+\hat{G})_{R}^{-1}\hat{N}(\lambda_{2})$
as its right inverse, we can equivalently rewrite the above equation
as 

\begin{equation}
\left(\hat{I}+\lambda_{1}(\hat{L}+\hat{G})_{R}^{-1}\hat{N}(\lambda_{2})\right)|V>=\hat{\Pi}(\lambda_{2})|V>\label{eq:28}
\end{equation}
 with projector 

\begin{eqnarray}
\hat{\Pi}(\lambda_{2})= & \hat{I}-(\hat{L}+\hat{G})_{R}^{-1}\hat{N}(\lambda_{2})\hat{N}_{l}^{-1}(\lambda_{2})(\hat{L}+\hat{G})\equiv\nonumber \\
 & \hat{I}-(\hat{L}+\hat{G})_{R}^{-1}\hat{Q}_{l}(\lambda_{2})(\hat{L}+\hat{G})\label{eq:29}
\end{eqnarray}
 In the last equation we have introduced the projector $\hat{Q}_{l}(\lambda_{2})=\hat{N}(\lambda_{2})\hat{N}_{l}^{-1}(\lambda_{2})$.
It is interesting to notice that similar equation as (\ref{eq:28})
we get if we multiply Eq.\ref{eq:25} by a right inverse operator
$(\hat{L}+\hat{G})_{R}^{-1}$:

\begin{eqnarray}
 & \left(\hat{I}+\lambda_{1}(\hat{L}+\hat{G})_{R}^{-1}\hat{N}(\lambda_{2})\right)|V>=\nonumber \\
 & \hat{P}_{L+G}|V>+(\hat{L}+\hat{G})_{R}^{-1}|0>_{info}\label{eq:30}
\end{eqnarray}
 with projector 

\begin{equation}
\hat{P}_{L+G}=\hat{I}-(\hat{L}+\hat{G})_{R}^{-1}(\hat{L}+\hat{G})\label{eq:31}
\end{equation}
 projecting on the null space of the operator $(\hat{L}+\hat{G})$.
Comparing equations (\ref{eq:28}) with (\ref{eq:30}) we get the
following equality upon undefined elements of the both equations:

\begin{equation}
\hat{\Pi}(\lambda_{2})|V>=\hat{P}_{L+G}|V>+(\hat{L}+\hat{G})_{R}^{-1}|0>_{info}\label{eq:32}
\end{equation}

\subsection{A three parameter theory and one-side invertible operators $\hat{N}(\lambda_{2})$}

We consider the operator

\begin{equation}
\hat{N}=\frac{\hat{N}(\lambda_{2})}{\hat{I}+\lambda_{3}\hat{N}(\lambda_{2})}\label{eq:33}
\end{equation}
 We now have an equation:

\begin{eqnarray}
 & \left(\hat{L}+\lambda_{1}\frac{\hat{N}(\lambda_{2})}{\hat{I}+\lambda_{3}\hat{N}(\lambda_{2})}+\hat{G}\right)|V>=\nonumber \\
 & \hat{P}_{0}\left(\hat{L}+\lambda_{1}\frac{\hat{N}(\lambda_{2})}{\hat{I}+\lambda_{3}\hat{N}(\lambda_{2})}+\hat{G}\right)|V>\equiv|0>_{info}\label{eq:34}
\end{eqnarray}
 with three parameters: the major coupling constant - $\lambda_{1}$,
the minor coupling constant - $\lambda_{2}$ and the regularization
parameter - $\lambda_{3}$, for which, if $\lambda_{3}=0$, then $\hat{N}=\hat{N}(\lambda_{2})$.
For the possible interpretation of these and other constants occurring
in the operators $\hat{L},\hat{N}$ and $\hat{G}$ see App.5.

From Eq.\ref{eq:34} 

\begin{eqnarray}
 & \left\{ \left(\hat{I}+\lambda_{3}\hat{N}(\lambda_{2})\right)\left(\hat{L}+\hat{G}\right)+\lambda_{1}\hat{N}(\lambda_{2})\right\} |V>=\nonumber \\
 & \left(\hat{I}+\lambda_{3}\hat{N}(\lambda_{2})\right)|0>_{info}\label{eq:35}
\end{eqnarray}

\subsection*{Left invertible operator $\hat{N}(\lambda_{2})$}

For a left invertible operator $\hat{N}(\lambda_{2})$, see (\ref{eq:35}),
we can transform this equation as follows: 

\begin{eqnarray}
 & \left\{ \left(\hat{N}_{l}^{-1}(\lambda_{2})+\lambda_{3}\hat{I}\right)\left(\hat{L}+\hat{G}\right)+\lambda_{1}\hat{I}\right\} |V>=\nonumber \\
 & \left(\hat{N}_{l}^{-1}(\lambda_{2})+\lambda_{3}\hat{I}\right)|0>_{info}\label{eq:36}
\end{eqnarray}
 For a large absolute value of the major coupling constant $\lambda_{1}$,
this equation can be solved by the perturbation method with the perturbation
operator $\lambda_{1}^{-1}\hat{N}_{l}^{-1}(\lambda_{2})\left(\hat{L}+\hat{G}\right)$.
In this case, the zeroth order approximation is approximately described
by the equation:

\begin{equation}
\left\{ \frac{\lambda_{3}}{\lambda_{1}}\left(\hat{L}+\hat{G}\right)+\hat{I}\right\} |V>=\frac{\lambda_{3}}{\lambda_{1}}|0>_{info}\label{eq:37}
\end{equation}
 where we have assumed that $\lambda_{1}\approx\lambda_{3}$. For
$\lambda_{1}>>>\lambda_{3}$, 

\begin{equation}
|V>\approx\frac{\lambda_{3}}{\lambda_{1}}|0>_{info}\label{eq:38}
\end{equation}
 In other words, in this case, in the zeroth order approximation,
no local information is contained in the generating vector |V>.

\subsection*{Right inverible operator $\hat{N}(\lambda_{2})$}

In this case there is an operator $\hat{N}_{R}^{-1}(\lambda_{2})$
for which

\begin{equation}
\hat{N}(\lambda_{2})\hat{N}_{R}^{-1}(\lambda_{2})=\hat{I}\label{eq:39}
\end{equation}
see Eq.\ref{eq:35}, and Eq.\ref{eq:34} can be transformed as follows:

\begin{eqnarray}
 & \left\{ \hat{I}+\left[\lambda_{3}\hat{N}(\lambda_{2})\left(\hat{L}+\hat{G}\right)\right]_{R}^{-1}\left(\hat{L}+\hat{G}+\lambda_{1}\hat{N}(\lambda_{2})\right)\right\} |V>=\nonumber \\
 & \hat{P}_{N}(\lambda_{2})|V>+\left[\lambda_{3}\hat{N}(\lambda_{2})\left(\hat{L}+\hat{G}\right)\right]_{R}^{-1}|0>_{info}\label{eq:40}
\end{eqnarray}
 with the projector 

\begin{eqnarray}
 & \hat{P}_{N}(\lambda_{2}) & =\hat{I}-\left[N(\lambda_{2})\left(\hat{L}+\hat{G}\right)\right]_{R}^{-1}\hat{N}(\lambda_{2})\left(\hat{L}+\hat{G}\right)\equiv\nonumber \\
 & \hat{I}- & \left(\hat{L}+\hat{G}\right)_{R}^{-1}\hat{Q}_{N}(\lambda_{2})\left(\hat{L}+\hat{G}\right)\label{eq:41}
\end{eqnarray}
 where the projector $\hat{Q}_{N}(\lambda_{2})=\hat{N}(\lambda_{2})_{R}^{-1}\hat{N}(\lambda_{2})$.
It turns out that for many polynomial operators $\hat{N}$, Eq.\ref{eq:40}
are closed, see \cite{Han 2010,Han 2010-1}.

\section{A few examples of right and left invertible operators}

We give here a few examples of operators $\hat{N}$ which are right
or left invertible operators and to which it is easy to construct
appropriate inverses. Moreover, such operators appear in varies field
theories. So, let us define

\begin{equation}
\hat{N}=\int d\tilde{x}\hat{\eta}^{\star}(\tilde{x})\hat{\eta}^{2}(\tilde{x})+\hat{P}_{0}\int d\tilde{x}\hat{\eta}(\tilde{x})f(\tilde{x})\label{eq:42}
\end{equation}
 A right inverse to this operator is 

\begin{equation}
\hat{N}_{R}^{-1}=1/2\int d\tilde{y}\hat{\eta}^{\star}(\tilde{y})^{2}\hat{\eta}(\tilde{y})+1/2\int d\tilde{y}\hat{\eta}^{\star}(\tilde{y})\label{eq:43}
\end{equation}
 if $\int d\tilde{x}f(\tilde{x})=2$. An example of \textbf{left invertible}
operator is

\begin{equation}
\hat{N}\equiv\hat{N}(\lambda_{2})=\int d\tilde{x}\hat{\eta}^{\star}(\tilde{x})H(\tilde{x})\left(M[\tilde{x};\hat{\eta};\lambda_{2}]\right)_{R}^{-1}\hat{I}\label{eq:44}
\end{equation}
 where $\left(M[\tilde{x};\hat{\eta};\lambda_{2}]\right)_{R}^{-1}$,
for every $\tilde{x}$ is a right, or both sides, inverse to the operator
$M[\tilde{x};\hat{\eta};\lambda_{2}]$ with a general operator-valued
function M. As a simple example of operator-valued function, let us
take 

\begin{equation}
M[\tilde{x};\hat{\eta};\lambda_{2}]=\hat{I}-\lambda_{2}\hat{\eta}(\tilde{x})\label{eq:45}
\end{equation}
 In the case of the right invertible operator (\ref{eq:45}) we have
a large selection of right inverses, see \cite{Przeworska 1988},
\cite{Han 2010}. Our choice can be inspired by the formal formula:

\begin{equation}
\left(\hat{I}-\lambda_{2}\hat{\eta}(\tilde{x})\right)^{-1}=\hat{I}+\sum_{n=1}^{\infty}\left(\lambda_{2}\hat{\eta}(\tilde{x})\right)^{n}=\left(\hat{I}-\lambda_{2}\hat{\eta}(\tilde{x})\right)_{R}^{-1}\label{eq:46}
\end{equation}
which, for a small absolute value of the minor coupling constant $\lambda_{2}$,
can be used to approximate the models such as the $\varphi^{3}$,
$\varphi^{4}$ etc. One can choose the formula (\ref{eq:46}) in such
a way that Eq. are closed, see \cite{Han 2010}, 

A left inverse to the operator $\hat{N}(\lambda_{2})$ is given by
formula:

\begin{equation}
\hat{N}_{l}^{-1}\equiv\hat{N}_{l}^{-1}(\lambda_{2})=\int d\tilde{y}E(\tilde{y})M[\tilde{y};\hat{\eta};\lambda_{2}]\hat{\eta}(\tilde{y})\label{eq:47}
\end{equation}
 with restriction $\int d\tilde{x}E(\tilde{x})H(\tilde{x})=1$. To
see the above statements we have to use the Cuntz relations (\ref{eq:6}).
It is also interesting that for a nonpolynomial operator (\ref{eq:44})
the variables $\tilde{x}$ can be continuous. For other choices of
the formula (\ref{eq:47}), see \cite{Han 2011-2}. See also \cite{Han 2008}.

\section{Equations with different operators $\hat{N}(\lambda_{2})$. A closure
principle}

We consider the model with the left invertible operator:

\begin{eqnarray}
 & \hat{N}(\lambda_{2})=\int d\tilde{x}\hat{\eta}^{\star}(\tilde{x})\left(\hat{I}+\lambda_{2}\hat{\eta}(\tilde{x})\right)+\lambda_{2}\hat{P}_{0}=\nonumber \\
 & \lambda_{2}\hat{I}+\int d\tilde{x}\hat{\eta}^{\star}(\tilde{x})\equiv\lambda_{2}\hat{I}+\hat{G}'\label{eq:48}
\end{eqnarray}
In the vector form, Eq.\ref{eq:9}, for n-pi, are:

\begin{equation}
\left(\hat{L}+\lambda_{1}\hat{N}(\lambda_{2})+\hat{G}\right)|V>=\hat{P}_{0}+\lambda_{2}\hat{P}_{0}=|0;\lambda_{2}>_{info}\label{eq:49}
\end{equation}
 or

\begin{equation}
\left\{ \left(\hat{L}+\lambda_{1}\lambda_{2}\hat{I}\right)+\left(\hat{G}+\lambda_{1}\hat{G}'\right)\right\} |V>=|0;\lambda_{2}>_{info}\label{eq:50}
\end{equation}
 These equations can be solved term by term because the equations
are not branching. Using however a left inverse operator 

\begin{equation}
\hat{N}_{l}^{-1}(\lambda_{2})=\int d\tilde{y}H(\tilde{y})\left(\hat{I}+\lambda_{2}\hat{\eta}(\tilde{y})\right)_{l}^{-1}\hat{\eta}(\tilde{y})\label{eq:51}
\end{equation}
 with $\int d\tilde{x}H(\tilde{x})=1$, multiplying Eq.\ref{eq:50}
by this operator, we get, for n-pi, highly branching system of equations:

\begin{equation}
\left\{ \hat{N}_{l}^{-1}(\lambda_{2})\left(\hat{L}+\hat{G}\right)+\lambda_{1}\hat{I}\right\} |V>=\hat{N}_{l}^{-1}(\lambda_{2})|0;\lambda_{2}>_{info}=0\label{eq:52}
\end{equation}
if the operator $\hat{N}_{l}^{-1}(\lambda_{2})$ in Eq.\ref{eq:52}
is related in some way to the formulas

\begin{equation}
\left(\hat{I}+\lambda_{2}\hat{\eta}(\tilde{y})\right)_{l}^{-1}\Leftrightarrow\left(\hat{I}+\lambda_{2}\hat{\eta}(\tilde{y})\right)^{-1}\Leftrightarrow\sum_{n=0}^{\infty}\left(-1\right)^{n}\left(\lambda_{2}\hat{\eta}(\tilde{y})\right)^{n}\label{eq:53}
\end{equation}
 The closure problem in the above model is artificially iniciated
by an inversion of the left invertible operator $\hat{N}(\lambda_{2})$. 

The model with left invertible operator:

\begin{equation}
\hat{N}(\lambda_{2})=\int d\tilde{x}\hat{\eta}^{\star}(\tilde{x})\left(\hat{I}+\lambda_{2}\hat{\eta}(\tilde{x})\right)_{R}^{-1}+\hat{P}_{0}\label{eq:54}
\end{equation}
 leads to closed equations for n-pi, if a right inverse operator appearing
in the formula (\ref{eq:54}) is defind as follws:

\begin{equation}
\left(\hat{I}+\lambda_{2}\hat{\eta}(\tilde{y})\right)_{R}^{-1}=\left(\hat{I}+(\lambda_{2}\hat{\eta}(\tilde{y}))_{R}^{-1}\right)^{-1}(\lambda_{2}\hat{\eta}(\tilde{y}))_{R}^{-1}\Leftrightarrow\left(\hat{I}+\lambda_{2}\hat{\eta}(\tilde{y})\right)^{-1}\label{eq:55}
\end{equation}
 where according to Cuntz relations (\ref{eq:6}),

\begin{equation}
(\lambda_{2}\hat{\eta}(\tilde{y}))_{R}^{-1}=\lambda_{2}^{-1}\hat{\eta}^{\star}(\tilde{y})\label{eq:56}
\end{equation}
(discrete case). In this case, the operator (\ref{eq:54}), 

\begin{equation}
\hat{N}(\lambda_{2})=\int d\tilde{x}\hat{\eta}^{\star}(\tilde{x})\frac{\lambda_{2}^{-1}\hat{\eta}^{\star}(\tilde{x})}{\hat{I}+\lambda_{2}^{-1}\hat{\eta}^{\star}(\tilde{x})}\label{eq:57}
\end{equation}
 and since the operators $\hat{\eta}^{\star}(\tilde{x})$ are lower
triangular operators, Eq.\ref{eq:9}: 

\[
\left(\hat{L}+\lambda_{1}\hat{N}(\lambda_{2})+\hat{G}\right)|V>=|0;\lambda_{2}>_{info}
\]
 is closed. Now, even for the operator $\hat{G}=0$, the vacuum $|0;\lambda_{2}>_{info}$
gives no vanishing contributions to the generating vector |V>, constructed
by, e.g., the formula (\ref{eq:19}). We can also get closed equations
for n-pi in the case of operators:

\begin{equation}
\hat{N}(\lambda_{2})=\int d\tilde{x}\hat{\eta}^{\star}(\tilde{x})\frac{\lambda_{2}^{-1}\hat{\eta}^{\star}(\tilde{x})}{\hat{I}+\lambda_{2}^{-1}\hat{\eta}^{\star}(\tilde{x})}\hat{\eta}(\tilde{x})^{k},\; k=1,2\label{eq:58}
\end{equation}
 which for $\lambda_{2}\ggg1$ are related formally with the unit
and $\varphi^{3}$ models. 

Models based on the operators:

\begin{equation}
\hat{N}(\lambda_{2})=\int d\tilde{x}\hat{\eta}^{\star}(\tilde{x})\left(\hat{I}+\lambda_{2}\hat{\eta}^{j}(\tilde{x})\right)_{R}^{-1}\hat{\eta}^{k}(\tilde{x})+\hat{P}_{0},\; k=1,...,j+1\label{eq:59}
\end{equation}
 admit a broader class of branching equations, if a similar formula
to \ref{eq:55} is used. We can say that highly uncertain resolvent
type of operators appearing in definitions of presented models, we
define by using some kind of a \textit{closure principle}, \cite{Han 2010-1}. 

The resulting freedom in defining the theory can be removed, at least
in the perturbation theory, by using the perturbation principle, see
\cite{Han 2010-1} and \cite{Han 2010,Han 2011-2,Han 2010-1}. See,
however, App.7.

\section{A solution of infinite system of branching equations. An expansion
in the inverse powers of the major coupling constant $\lambda_{1}$.}

We consider now the model: 

\begin{equation}
\hat{N}(\lambda_{2})=\int d\tilde{y}H(\tilde{y})\left(\hat{I}+\lambda_{2}\hat{\eta}(\tilde{y})\right)_{l}^{-1}\hat{\eta}(\tilde{y})\label{eq:60}
\end{equation}
 which is a \textbf{right invertible }operator with a right inverse:

\begin{equation}
\hat{N}_{R}^{-1}(\lambda_{2})=\int d\tilde{x}\hat{\eta}^{\star}(\tilde{x})\left(\hat{I}+\lambda_{2}\hat{\eta}(\tilde{x})\right)\label{eq:61}
\end{equation}
 In this case, in equations for n-pi, (\ref{eq:9}), integrations
or summations appear, like in equations of statistical mechanics,
see \cite{Balescu 1978}. Eq.\ref{eq:9}

\begin{eqnarray}
 & \left(\hat{L}+\lambda_{1}\hat{N}(\lambda_{2})+\hat{G}\right)|V>=\nonumber \\
 & \hat{P}_{0}\left(\hat{L}+\lambda_{1}\hat{N}(\lambda_{2})+\hat{G}\right)|V>\equiv|0>_{info} & 62\label{eq:62}
\end{eqnarray}
is an infinite system of branching equations if the expansion (\ref{eq:53})
is somehow justified. Multiplying Eq.\ref{eq:62} by a right inverse,
(\ref{eq:61}), we get an equivalent equation

\begin{eqnarray}
 & \left\{ \lambda_{1}^{-1}\hat{N}_{R}^{-1}(\lambda_{2})\left(\hat{L}+\hat{G}\right)+\hat{I}\right\} |V>=\nonumber \\
 & \lambda_{1}^{-1}\hat{N}_{R}^{-1}(\lambda_{2})|0>_{info}+\hat{P}_{N}(\lambda_{2})|V>\label{eq:63}
\end{eqnarray}
 where the projector $\hat{P}_{N}(\lambda_{2})=\hat{I}-\hat{N}_{R}^{-1}(\lambda_{2})\hat{N}(\lambda_{2})$
projects on the null space of the operator $\hat{N}(\lambda_{2})$.
The vector projection, $\hat{P}_{N}(\lambda_{2})|V>$, is an arbitrary
element of the general solution to Eq.\ref{eq:62}. It can be identified
by means of the perturbation theory, with the perturbation parameter
$\lambda_{2}$, by means of which the unknown vector |V> is expanded
as:

\begin{equation}
|V>\equiv|V(\lambda_{2})>=\sum\lambda_{2}^{n}|V>^{(n)}\label{eq:64}
\end{equation}
 It is interesting to notice that in this model, the n-th approximation
$|V>^{(n)}$ , after inversion of the diagonal operator $\left(\int d\tilde{x}\hat{\eta}^{\star}(\tilde{x})\int d\tilde{y}H(\tilde{y})\hat{\eta}(\tilde{y})\right)$,
is expressed by the lower order approximations at the arbitrary value
of the coupling constant $\lambda_{1}$.

\section{Final remarks; Ockham's razor, vanishing of demarcation line between
$\hat{L}$ and $\hat{N}$ operators}

The presented work shows that in classical theories with some information
losses caused by smoothing data, the local information vacuum described
by the vector $|0>_{info}$ exerts similar effects on the process
of calculation as the quantum vacuum in the perturbation theory: Particularly,
in presence of external fields, $\hat{G}\neq\hat{0}$, phenomena would
go differently if the vector $|0>_{info}$would not occurred in the
formulas (\ref{eq:19}), (\ref{eq:32}), or (\ref{eq:38}). Moreover,
for a particular form of the operators $\hat{N}$, connected with
the so called non-polynomial nonlinearities, even at $\hat{G}=0$,
the classical vacuum plays an important role, if we take into account
possible indeterminacy of \textbf{operator-valued functions} and define
them as in Secs 4 -5, see also \cite{Han 2010-1} and \cite{Han 2010}.
In fact, we can say that derived equations are one of the ways of
the definition of such functions.

We recall here that appearance of this vector was the result of complement
of equations on the n-pi with the zero component in such a way that
the operators appearing in the equations become at least one-side
invertible. 

A more fundamental vector in FFS is the vector |0> by means of which
(local) n-pi about an arbitrary system is created. It looks as if
|0> contains information about the whole Universe. In this interpretation
\textit{intriguing is} that the vector $|0>_{info}$ describing the
n-pi (quasi-local info) and the vector $|0>$, describing Total information
(Global Info) about the Universe, are proportional to each other.
It looks as if the FFS have contained a metaphysical inspirations! 

By introducing of additional parameters, new insights into the closure
problem is also obtained. In\textit{ Sec.5} it is shown how easy closed
equations can be converted into infinite system of branching equations,
which are difficult to solve. On the other hand, by introducing the
regularization parameter, $\lambda_{3}$, the closed equations can
be derived for the large class of branching (not closed) equations,
for n-pi, Sec.3.3. The problem that then arises is the emergence of
undetermined projected vector $\hat{P}_{N}(\lambda_{2})|V>$, see
Eqs (\ref{eq:40}) and (\ref{eq:63}), which belongs to a larger subspace.
Derived relation (\ref{eq:32}) and the principle of perturbation
suggest a choice, see also \cite{Barnaby 2008}. Of course, in all
formulas, like in (\ref{eq:19}), the symmetry condition (\ref{eq:17})
can be used. 

The lesson that seems to result from this and previous works, especially
with \cite{Han 2010-1}, is that from only the standpoint of equations
and approximations, it is more useful to operate with a more complex
nonpolynomial interactions, which approximate the more simple polynomial
interactions. Could be it a sign of the emergence of a new paradigm?
The analogy that here arises to me is similar to replacing of one
computer - with many - working together (cloud computation). Otherwise,
nonlinearities, which appear at the lower level of description, need
not to have polynomial forms. Nevertheless, we should still keep in
mind \textbf{Ockham's rezor} law called also law of economy or law
of parsimony that - Pluralitas non est ponenda sine necessitate; 'Plurality
should not be posited without necessity'. The principle gives precedence
to simplicity; of two competing theories, the simplest explanation
of an entity is to be preferred. The principle is also expressed 'Entities
are not to be multiplied beyond necessity.' In the present work, this
principle is implemented in Sec.3 by using the same operators for
different values of parameters $\lambda$. 

There is another noteworthy fact, namely, using only formulas with
one-sided reversibility. The fact that such property of operators
is associated with the fundamental premise of set theory, namely the
Axiom of Choice, raises hope that the derived formulas are not just
formal expressions.

Another important feature of this and previous papers is blurring
the differences between the operators $\hat{L}$ and $\hat{N}$. This
is manifested in the fact that in the transformation equations and
expansions inverse operators to both of these operators occur. Somewhat
similar phenomenon has been present in the very interesting \textit{homotopy
analysis method}, see e.g. \cite{Ajou 2010}. In this method the \textit{embedding
parameter} $q\in[0,1]$, or rather the \textit{weighting factor} of
relevancy of operators $\hat{L}$ and $\hat{N}$ , is introduced to
Eq.\ref{eq:9} - to get

\begin{eqnarray}
 & \left((1-q)\hat{L}+\lambda q\hat{N}+\frac{1}{2}\hat{G}\right)|V>=\nonumber \\
 & \hat{P}_{0}(1-q)\hat{L}|V>+\lambda q\hat{P}_{0}\hat{N}|V>+\frac{1}{2}\hat{P}_{0}\hat{G}\equiv|0;q>_{info}\label{eq:65}
\end{eqnarray}
 For $q=\frac{1}{2}$ , the original Eq.\ref{eq:9} is obtained. In
this case, we can use, for example, the \textbf{expansion with respect
to the weighting factor q}:

\begin{equation}
|V>=\sum_{j}q^{j}|V>^{(j)}\label{eq:66}
\end{equation}
in which the j-th order terms, $|V>^{(j)}$, are expressed by the
(j-1)-th. In such formulae the both operators, $\hat{L}$and $\hat{N}$,
enter in a similar way. You can see what it gives, when Eq.\ref{eq:65},
with the expansion parameter $q$, we transform as we did with the
Eq.\ref{eq:9}, with the expansion parameter $\lambda$. In the case
of Eq.\ref{eq:65}, we get, for example:

\begin{eqnarray}
 & \left(\hat{I}+\lambda\frac{q}{1-q}\hat{L}_{R}^{-1}\hat{N}+\frac{1}{1-q}\hat{L}_{R}^{-1}\hat{G}\right)|V>=\nonumber \\
 & \frac{1}{1-q}\hat{L}_{R}^{-1}|0;q>_{info}+\hat{P}_{L}|V>\label{eq:67}
\end{eqnarray}
 In this way the parameter q should enter the perturbative theory,
if we want the operators $\hat{L}$ and $\hat{N}$ in the Eq.\ref{eq:65}
were included approximately with the same weight, for $q\approx1/2$.
And this is a sence of the homotopy approach. 

Another similarity in treatment of the operators $\hat{L}$ and $\hat{N}$
is that |V>, for $q=\{0,1\}$, satisfies the following equations:

\begin{equation}
\left(\hat{L}+\frac{1}{2}\hat{G}\right)|V>=\hat{P}_{0}\hat{L}|V>+\frac{1}{2}\hat{P}_{0}\hat{G}\label{eq:68}
\end{equation}
and 

\begin{equation}
\left(\lambda\hat{N}+\frac{1}{2}\hat{G}\right)|V>=\lambda\hat{P}_{0}\hat{N}|V>+\frac{1}{2}\hat{P}_{0}\hat{G}\label{eq:69}
\end{equation}
and perhaps, 'this is it'! I believe that this or similar treatment
of operators $\hat{L}$ and $\hat{N}$ in the proposed approximations
to the vector |V> extend their effectivness. For example, in the case
of Eq.\ref{eq:9} we get the expansion (\ref{eq:65}) in which the
expansion parameter, $q=1/2<1$, corresponds to the original theory
(\ref{eq:9}). Another example is the closed equations derived in
Sec.3.3, and so on. We can say that due to vanishing of the demarcation
line between operators $\hat{L}$ and $\hat{N}$, new possibilities
are open. Also by giving the same weight of importancy of operators
L and N perhaps it is in some sense an implementation of the Ockham's
razor law.

Now I would like to write a few sentences about the opposite phenomenon,
namely the desirability of the presence of demarcation lines in a
certain areas of the theory or its equations. This is when we want
to take advantage of their various properties such as algebraic, geometric,
topological, etc., in order to achieve satisfactory solutions. I believe
that this type of strategy is made possible by using the FFS and n-pi:)

\section{App.1 about equations, components of the generating vector |V> and
smoothed and nonsmoothed solutions to Eq.71}

We assume that these components have the following interpretation
expressed by the equalities:

\begin{equation}
V(\tilde{x}_{(n)})=<\varphi(\tilde{x}_{1})\cdots\varphi(\tilde{x}_{n})>\label{eq:70}
\end{equation}
 where the field $\varphi(\tilde{x})$ is a general solution of the
'micro'or local equation:

\begin{equation}
L[\tilde{x};\varphi]+\lambda N[\tilde{x};\varphi]+G(\tilde{x})=0\label{eq:71}
\end{equation}
 with linear (L), nonlinear (N) and external (G) parts. The symbol
<> means operation of averaging or smoothing of solutions to Eq.\ref{eq:71}. 

<...> can also mean what does the mind during our vision and in general
to survive, see App.4. 

We have

\begin{equation}
\varphi(\tilde{x})=\varphi[\tilde{x};\alpha]\label{eq:72}
\end{equation}
 where $\alpha$denots all the boundary and initial conditions related
to Eq.\ref{eq:71}. In our case, these are \textit{'random'} variables.
So, with a probability density or a weight density, we get

\begin{equation}
<\varphi(\tilde{x}_{1})\cdots\varphi(\tilde{x}_{n})>=\int\delta\alpha\varphi[\tilde{x}_{1};\alpha]\cdots\varphi[\tilde{x}_{n};\alpha]W[\alpha]\label{eq:73}
\end{equation}
 were the symbol $\delta\alpha$ means the functional integration
(ensemble or statistical smoothing). 

There is also another possible way of smoothing the field $\varphi$
by using one or a few multiple integrals. It is so called the moving
average, also called the rolling average. In this case we have:

\begin{equation}
<\varphi(\tilde{x}_{1})\cdots\varphi(\tilde{x}_{n})>=\int dw\varphi[\tilde{x}_{1};\alpha_{w}]\cdots\varphi[\tilde{x}_{n};\alpha_{w}]W(w)\label{eq:74}
\end{equation}
where $\alpha_{w}(\tilde{x})=\alpha(\tilde{x}-\tilde{w})$, see also
App.6. Naturally, if Eq.\ref{eq:71} is not translationally invariant,
we can not use the formula (\ref{eq:74}), if we hope to get the same
Eq.\ref{eq:9}. In General Relativity, there is additional difficulty
related to the formula (\ref{eq:74}) caused by the fact that you
can not add tensors defined at different spacetime points (averaging
problem), see e.g., \cite{Wilt 2011}. In the case of (\ref{eq:73})
we make averaging/smoothing of tensors selected at the same points. 

In both cases, e.g., for 

\begin{equation}
W(w)=\delta(w)\label{eq:75}
\end{equation}
the formalism presented can be used, at least for certain nonlinearities,
$N[\tilde{x};\varphi]$, to construct nonaveraged solutions to Eq.\ref{eq:71}.
In the case of Eq.\ref{eq:73} we can use the weight density

\begin{equation}
W[\alpha]=\delta[\alpha-\alpha_{0}]+W'[\alpha]\label{eq:76}
\end{equation}
with the functional Dirac's $\delta$ and a smooth nonsingular functonal
$W'[\alpha]$. Then, the whole set of vectors $\{|V;W',\alpha_{0}>\}_{W'\subset smooth\, functionals}$
can be constructed. With the help of different $\alpha_{0}$, the
whole \textbf{class} (collection, family) of such \textbf{sets} can
also be constructed. From AC a set of vectors $|V>$constituated from
exact solutions to Eq.\ref{eq:71} and their products exists. This
statement can have non-trivial value if the existence of particular
sets of the collection would be an easier task to prove than separate
solutions. 

It is also interesting to notice that smoothing/averaging operation
<...> , at fixed constatnts of motion of Eq.\ref{eq:9}, do not change
Eq.\ref{eq:71} even so it correspons to reduction of the number of
freedom.

\section{App.2 about the Axiom of Choice (AC)}

If we define a choice function: \textit{a choice function on a class
H of nonempty sets }is a map f with domain H such that $f(X)\in X$
for every $X\in H$, then AC can be formulated as:

'Any class of nonempty sets has a choice function.

Finally AC is easily shown to be equivalent (in the usual set theories)
to: 

Any surjective function has a right inverse.' (Stanford Encyclopedia
of Philosophy (Internet)

xxxxx

Generalisation:

'A class is a collection of sets: for any property, we can form a
class of all sets with property P. But there is no surjection from
a set to a class that itself is not a set. Every set is a class. A
conglomerate is collection of classes: for any property P, we can
form a conglomerate of classes with property P. Moreover, we assume
an AC for Conglomerates: for each surjection $f:\: X\rightarrow Y$
of conglomerates, there exists an injection $g:\: Y\rightarrow X$
with $f\odot g=idY$, the identity of $Y$. Every class is also a
conglomerate.' (\cite{Wagh 2006}). 

xxxxx

Comments:

App.2 means also that right invertible operators (surjective functions)
considered in the paper can transform the FFS X in another FFS Y. 

From AC also results that if we have a relation defined as 

\begin{equation}
x\sim y\quad iff\quad f(x)=f(y)
\end{equation}
 then $g\odot f$ (a projector) maps everything in each equivalence
class to one of its members.

\section{App.3 about complex and complicated systems. A consilience in action }

'The distinction made is that a \textit{complicated system} is composed
of \textbf{many parts}, which can be seen as independent of the other
parts. Sort of like in a car, if you remove the front seat, the car
can still run and the general functionality of the car is left unaffected.
But in a \textit{complex system}, the components are interconnected
in a way such that affecting on member will in turn affect the functionality
of all members contained within the system. In a car, removing the
engine belt would in turn stop the car from functioning.'

quotation from http://cuzzopaint.in/post/5012868509/complexity-complication.

The above definitions of complex and complicated systems are used
in economy and may be better understood in terms of control: 'The
\textbf{complex phenomena can be controlled because they are reducible
to a simpler description, so that the seemingly different events can
be treated similarly}'. This is quotation from \cite{Full 2007}. 

If the system is described by continuous or almost continuous field,
you could probably say that it consists of many parts. If these parts
are somehow interconnected, as in our case by Eq.\ref{eq:71}, one
can speak of a complex system. If not, and this is often related to
the linear term (see Quantum Field Theory), we can speak about a complicated
system. In practical economics, it is not possible to clearly separate
complexity from comlication (Ludvig von Mises, Friedrich Hayek), \cite{Full 2007}.
In the theory of evolution, the situation is similar and it is conceivable
that complication may reveal complexity, \cite{Full 2007}. See also
\cite{Burn 2011} for 'consilient' or 'vertically integrated' approach
to the study. 

It turns out that without definite knowledge concerning Eq.\ref{eq:71}
one can say something a bit about its averaged or smoothed solutions,
which, in any case of non-linear theory, are not solutions to that
equation. In the introduction of previous work, \cite{Han 2011-2},
we mentioned the possibility of using this approach in a wide range
of human activity. If the field $\varphi$ describes positioning of
letters of a novel and operation $<...>$ some kind of summarizing
of text fragments, we can hope that higher order of n-pi can describe
meaning of words or sentences. Even the reading is characterized by
the property which is common feature of many mathematical operations:
It strongly depends on the direction of the reading. It is not excluded
that the multi-level description of the mind and body, see \cite{Davies 1983},
will someday be reflected in the considered here equations for the
generating vector |V>.  After a cursory reading of the paper \cite{Wilson 1998},
our optimism is even greater! 

Another example of consilience phenomenon is relation between memory
and continuity. Greater precision 'in thinking about thinking' is
also the result of 'consilience in action' in the case of computer
science, linguistics, cybernetics and psychology, see \cite{Davies 1983}. 

At the end, let me quotate Jerome Iglowitz's paper (The Mind-Brain
Problem: An Introduction for Beginners): '{[}Hilbert's{]} revolution
lay in the stipulation that the basic or primitive concepts are to
be defined just by the fact that they satisfy axioms...{[}They{]}
acquire meaning only by virtue of the axiom system, and possess only
the content that it bestows upon them. They stand for entities whose
whole being is to be bearers of the relations laid down by the system'.
And one more quotation from the same paper this time by the Hilbert:
'If one is looking for other definitions of A 'point', e.g. through
paraphrase in terms of extensionless, etc., then I must indeed oppose
such attemps in the most decisive way; one is looking for something
one can never find because there is nothing there; and everything
gets lost and becomes vague and tangled and and degenerates into a
game of hide and seek.' 

It is not excluded that the sentence: 'there is no mind' resulting
from the above quotations is not purelly materialistic but it is only
expression of very complex phenomena which we call the mind.

\section{App.4 about a rejection and the mind}

'Vision is not a simple reflection of (shooting) environment. It involves
the continuous exploration and \textbf{rejection} of what the mind
considers it unnecessary.' from \cite{Przyb 2011}. 

It is believed that the selection process led to the survival of the
minds, which especially effectively were avoiding superfluous information,
see also \cite{=00017Burek 1990}. In this way a person acquired the
so-called opportunity to act (affordances), see J.J. Gibson and M.
B\l{}aszak in \cite{=00017Burek 1990}. Getting rid of the huge amount
of information overload would require large amounts of energy which
would have a negative impact on reproduction and survival chances
of organisms.

\section{App.5 about a possible interpretation of explicitly appearing constants
in Eq.\ref{eq:71}. A more fundamental equations?}

Eq.\ref{eq:9} can be also used in the case when constants appearing
in this equation or Eq.\ref{eq:71} are 'random' quantities. In such
cases we have to solve first Eq.\ref{eq:9} and after that we can
use the averaging operation <...> now with respect to these constatants.
It would be interesting to investigate wether all these constants
can be treated as values of corresponding integrals of motion or constants
of motion of an appropriately generalized Eq.\ref{eq:71}. It would
be also interesting to find whether such constants of motion are related
to some symmetries like in Noether's theorem (1918), see \cite{Kozlov 1995}.
In this way, constants, especially the fundamental constants are treated
as indications that in the background of Eq.\ref{eq:71} lie in a
more simple equation. This belief stems from the fact that taking
into account in Eq.\ref{eq:71} the available in sequence - constants
of motion - leads to increasing complexity of equations to a smaller
number of degrees of freedom. Particularly desirable would be an equation
in which the role of the eliminated degrees of freedom would amount
to maintain the stability of fundamental constants. It is also possible
that these additional variables could be used to describe the unexplained
phenomena of the dark material world!?

By the way, I note that in the case of equations with the constants
of motion due to the corresponding symmetry, it is easy to implement
the inverse program outlined above. see in particular the end of App.1.

\section{App.6 about smoothing or averaging the linear Eq.(\ref{eq:71}).
Ergodic case.}

As we know from this or other papers, the freedom of a given theory
is usually reduced to its linear part. So, it is useful to discuss
the case of linear Eq.\ref{eq:71}. From that point of view, we assume
that equations are linear and homogeneous. In this case we denote
the general solution to such equations as 

\begin{equation}
\varphi(\tilde{x})=\varphi_{0}[\tilde{x},\alpha]\label{eq:78}
\end{equation}
 Due to the superposition principle satisfied by solutions of any
linear equation, if $\varphi_{0}[\tilde{x},\alpha_{i}]$ denote varies
solutions of the linear Eq.\ref{eq:71} satisfying varies additional
conditions $\alpha_{i}$, then their arbitrary combination:

\begin{equation}
\sum_{i}c_{i}\varphi_{0}[\tilde{x},\alpha_{i}]\equiv\varphi_{0}[\tilde{x},\alpha]\label{eq:79}
\end{equation}
 is also a solution of the same linear Eq.\ref{eq:71}. The same can
be said of the linear combinations of functions:

\begin{equation}
\sum_{i}c_{i}\varphi_{0}[\tilde{x}+\tilde{w}_{i},\alpha]\equiv\varphi_{0}[\tilde{x},\alpha']\label{eq:80}
\end{equation}
where the additional condition (AD), $\alpha,\alpha'$ in Eqs\ref{eq:79},\ref{eq:80}
also depend on the set of coefficients $c_{i}$ The linear combinations
of solutions in the sum (\ref{eq:79}, \ref{eq:80}) can be treated
as prototypes of corresponding averages'' discussed in App.1. Ergodicity
means that these two types of averages should be equal to each other.
Of course, we can only hope that\textbf{\textit{ asymptotic ergodicity}}
takes place, which here means that \textit{asymtotes of solutions
are independent of initial conditions}. We mean by this that, for
sufficiently large time, the values of solutions can be found in a
particular set. 

Of course, if nonlinearity is taken into account in Eq.\ref{eq:71},
then in averaged/smoothed solutions described by Eq.\ref{eq:9} a
dependence on the method of averaging/smoothing can be seen.

For an excellent illustration of the ergodicty idea also in the social
sciences, see Wikipedia: What Is Ergodicity?''.

Can we inferre from the above properties of linear equations that
if solutions oscillate around some regular functions, either of these
functions or other regular functions are also solutions to these linear
equations? It seems that averaging or smoothing processes of solutions
of a linear equation lead to more smooth solutions of the same equation
if we consider linear equations with constant coefficients. In this
case both kind of averaging/smoothing can be executed. The case of
linear equations with variable coefficients does not allow to make
averages with the help of formula (\ref{eq:80}). Such linear equations
(\ref{eq:71}) (with $\hat{N}=0$) sometimes are associated with nonlinear
equations, see App.7. In such cases, it is not excluded that $<\varphi(\hat{x})>$
is no more regular than $\varphi[\hat{x};\alpha]$, for a certain
set of additional conditions $\alpha$ , but agregated'' quantity
$<\varphi(\hat{x})>$ better reflects human capabilities.

\section{App.7 Nonlinear expressed by linear equation solutions}

The Riccati equations are simple example of a situation in which the
linear equations are related, by certain change of variables, to the
nonlinear equations. The general case of such situation, expressed
in terms of solutions, can be described by the following functional
equation:

\begin{equation}
\varphi[\hat{x};\alpha]=F[\hat{x};\psi[\bullet;\alpha]]\label{eq:81}
\end{equation}
 in which the functions $\psi[\hat{y};\alpha]$ satisfies a linear
equation. We see that now all n-pi related to the fields $\varphi[\hat{x};\alpha]$
can be constructed by means of the n-pi related to the fields $\psi[\hat{y};\alpha]$,
if the Volterra series:

\begin{equation}
\varphi[\hat{x};\alpha]=F[\hat{x};\psi[\bullet;\alpha]]=\sum_{n}\int d\hat{x}_{(n)}F(\hat{x};\hat{x}_{(n)})\psi[\hat{x}_{1};\alpha]\cdots\psi[\hat{x}_{n};\alpha]\label{eq:82}
\end{equation}
 is used. We can get such series if Eq.\ref{eq:71} is solved by using
a perturbation theory. We must point out here that the relation (\ref{eq:82})
depends on the way in which the perturbation parameter $q$ is introduced,
see Eq.\ref{eq:65}. We must also say that the similar relation, which
connects a solution of nonlinear equation solution (NES) with linear
(LES) in the case of Riccati equations:

\begin{equation}
NES=NES_{1}+\frac{1}{LES}\label{eq:83}
\end{equation}
where $NES_{1}$- denotes a particular solution of the nonlinear Ricci's
equation, is a warning against too great deal of trust in relationships
(\ref{eq:82}). The most surprising is reverse situation expressed
by the equation:

\begin{equation}
LES=\frac{1}{NES-NES_{1}}\label{eq:84}
\end{equation}
 which admits the expansion (\ref{eq:82}) of LES with respect to
NES! That would be yet another argument for seeking unperturbative
approaches.

\end{document}